\documentclass{rnaastex}

\begin{document}

\title{Gaia confirms that SDSS\,J102915+172927 is a dwarf star}

\correspondingauthor{P. Bonifacio}
\email{Piercarlo.Bonifacio@observatoiredeparis.psl.eu}

\author{P. Bonifacio}
\affiliation{GEPI, Observatoire de Paris, Universit\'e PSL, CNRS, Place Jules
Janssen, F-92195 Meudon, France}

\author{E. Caffau}
\affiliation{GEPI, Observatoire de Paris, Universit\'e PSL, CNRS, Place Jules
Janssen, F-92195 Meudon, France}

\author{M. Spite}
\affiliation{GEPI, Observatoire de Paris, Universit\'e PSL, CNRS, Place Jules
Janssen, F-92195 Meudon, France}
\author{F.  Spite}
\affiliation{GEPI, Observatoire de Paris, Universit\'e PSL, CNRS, Place Jules
Janssen, F-92195 Meudon, France}

\author{P. Fran\c cois}
\affiliation{GEPI, Observatoire de Paris, Universit\'e PSL, CNRS, Place Jules
Janssen, F-92195 Meudon, France}
\author{S. Zaggia}
\affiliation{Istituto Nazionale di Astrofisica,
Osservatorio Astronomico di Padova, Vicolo dell'Osservatorio 5, 35122 Padova, Italy}  
\author{F. Arenou}
\affiliation{GEPI, Observatoire de Paris, Universit\'e PSL, CNRS, Place Jules
Janssen, F-92195 Meudon, France}
\author{R. Haigron}
\affiliation{GEPI, Observatoire de Paris, Universit\'e PSL, CNRS, Place Jules
Janssen, F-92195 Meudon, France}
\author{N. Leclerc}
\affiliation{GEPI, Observatoire de Paris, Universit\'e PSL, CNRS, Place Jules
Janssen, F-92195 Meudon, France}
\author{O. Marchal}
\affiliation{GEPI, Observatoire de Paris, Universit\'e PSL, CNRS, Place Jules
Janssen, F-92195 Meudon, France}
\author{P. Panuzzo}
\affiliation{GEPI, Observatoire de Paris, Universit\'e PSL, CNRS, Place Jules
Janssen, F-92195 Meudon, France}
\author{G. Plum}
\affiliation{GEPI, Observatoire de Paris, Universit\'e PSL, CNRS, Place Jules
Janssen, F-92195 Meudon, France}
\author{P. Sartoretti}
\affiliation{GEPI, Observatoire de Paris, Universit\'e PSL, CNRS, Place Jules
Janssen, F-92195 Meudon, France}


\section{} 

With a global metallicity of $Z \le 6.9\times 10^{-7}$ \citep{Caffau11,Caffau12} star
 SDSS\,J102915+172927 is, at the time of writing, the most metal-poor object known.
This star is also remarkable because it has  abundances of C and O
that are low enough that the metal line cooling proposed by
\citet{brom_loeb} cannot be efficient to allow a low-mass star 
to form. The very existence of this star has been taken as proof that 
other mechanisms, such as dust cooling \citep{schneider} and
fragmentation \citep{klessen} can allow to form low-mass stars
at metallicities below what is allowed by metal line cooling alone. 
 
Another of the distinctive features of  this star is that lithium has not been 
detected. 
This is at variance with what is observed in 
warm metal-poor stars of higher metallicity, which display a constant lithium abundance, 
the Spite plateau
\citep{Spite82}. 
As discussed in \citet{toposIV} amongst all the known
unevolved stars with [Fe/H]$< -4.1$ only two have a measured lithium
abundance and in both cases it is more than 0.2\,dex below the Spite plateau.
The effective temperature of  SDSS\,J102915+172927 
is 5811\,K \citep{Caffau12} and for such stars  
stellar evolution models do not predict extensive Li depletion, as
is the case for cooler stars \citep[see e.g.][]{SW01}. 
\citet{toposIV} suggested that  the effective temperature
at which convection begins to effectively deplete lithium
increases with decreasing [Fe/H].

\citet{MacDonald} have argued that the upper limit on the lithium abundance in 
 SDSS\,J102915+172927 is stringent enough that it  rules out an identification as  
a dwarf star. For this reason they argued that it is indeed a subgiant star. 
In the scenario in which the star is a dwarf, they estimate a distance of
$1.35\pm 0.16$ kpc, while in the subgiant scenario they estimate a distance
of $6.2\pm0.5$ kpc. 
The subgiant scenario has an important corollary: it allows for the star
to be formed in a C-enriched medium, where metal line cooling is effective, 
that has been subsequently enriched by a supernova of type Ia, that produced
the low C/Fe ratio observed.  

\begin{figure}[h!]
\begin{center}
\includegraphics[clip=true]{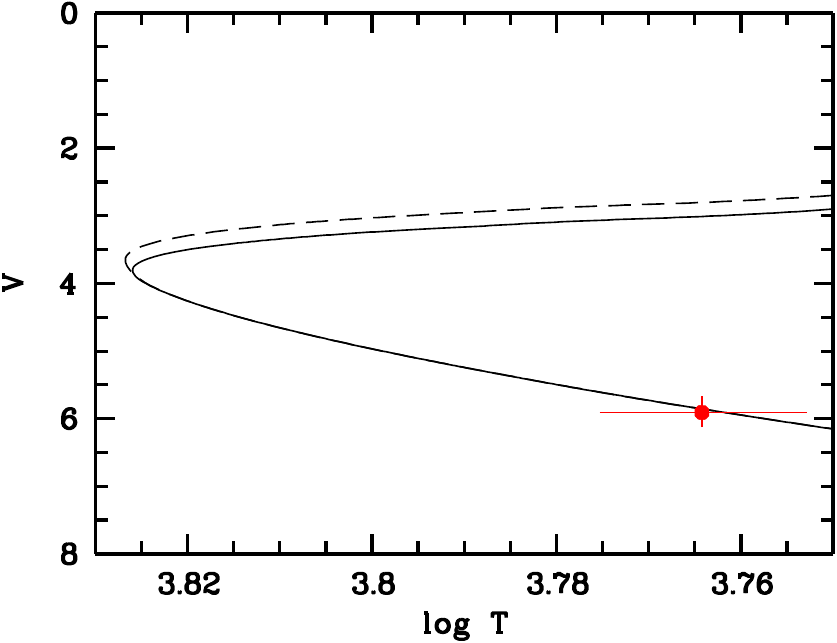}
\caption{The absolute magnitude of SDSS\,J102915+172927 derived
from the Gaia parallax and the Gaia photometry implies it is a dwarf
star. For comparison two isochrones (A. Chieffi, private communication)
computed with the FRANEC code 
for 14 Gyrs and $Z=2\times 10^{-6}$ (dashed line),
$Z=2\times 10^{-5}$ (solid line)
and $Z=2\times 10^{-4}$ (dotted line) are shown.\label{fig:1}}
\end{center}
\end{figure}

It is very difficult to strongly constrain the luminosity of
a warm low metallicity star from its spectrum. 
In the case of  SDSS\,J102915+172927, for example, we do not observe
any Fe{\sc ii} line, that could constrain the luminosity, by imposing 
the ionization equilibrium. 
The broad photometric comparison with theoretical isochrones of 
\citet{Caffau12} 
gave the dwarf as best solution with the subgiant scenario slightly less
favourable.
The Gaia mission \citep{Gaia} with its second data release
\citep{DR2,DR2_valid} has provided us a very accurate parallax
of $0.734\pm0.073$ mas. 
Thus the distance is $1.37^{+0.15}_{-0.12}$ kpc, in perfect
agreeement with what estimated by
\citet[][$1.37\pm0.20$ kpc]{Caffau12}.
Gaia also provides photometry for this
star $G=16.548$ and a colour $BP-RP=0.799$.
The star is only slightly reddened, the 
reddening maps of \citet{redmap} imply $E(B-V)=0.03$ mag.
We used the above information to  compute
the absolute V magnitude using the transformations 
provided by the Gaia DR2 documentation
(\url{https://gea.esac.esa.int/archive/documentation/GDR2/Data_processing/chap_cu5pho/sec_cu5pho_calibr/ssec_cu5pho_PhotTransf.html}).
This allows us to compare it to the metal-poor isochrones
computed by A. Chieffi (private communication) and shown in
the figure.
We can unambiguously conclude that 
 SDSS\,J102915+172927 is a dwarf. The subgiant scenario
proposed by \citet{MacDonald} can be definitely ruled
out, together with the hypothesis that  SDSS\,J102915+172927
was formed from a medium dominantly enriched by SNIa.
This reinforces the need for dust cooling and fragmentation
to form a star like this.

\acknowledgments
This work has made use of data from the European Space Agency (ESA) mission
{\it Gaia} (\url{https://www.cosmos.esa.int/gaia}), processed by the {\it Gaia}
Data Processing and Analysis Consortium (DPAC,
\url{https://www.cosmos.esa.int/web/gaia/dpac/consortium}). Funding for the DPAC
has been provided by national institutions, in particular the institutions
participating in the {\it Gaia} Multilateral Agreement.

\end{document}